# Enabling large-scale viscoelastic calculations via neural network acceleration

Phoebe M. R. DeVries, T. Ben Thompson, and Brendan J. Meade

Department of Earth and Planetary Sciences

Harvard University

Cambridge, Massachusetts, 02138

**Abstract.** One of the most significant challenges involved in efforts to understand the effects of repeated earthquake cycle activity are the computational costs of large-scale viscoelastic earthquake cycle models. Computationally intensive viscoelastic codes must be evaluated at thousands of times and locations, and as a result, studies tend to adopt a few fixed rheological structures and model geometries, and examine the predicted time-dependent deformation over short (<10 yr) time periods at a given depth after a large earthquake. Training a deep neural network to learn a computationally efficient representation of viscoelastic solutions, at any time, location, and for a large range of rheological structures, allows these calculations to be done quickly and reliably, with high spatial and temporal resolution. We demonstrate that this machine learning approach accelerates viscoelastic calculations by more than 50,000%. This magnitude of acceleration will enable the modeling of geometrically complex faults over thousands of earthquake cycles across wider ranges of model parameters and at larger spatial and temporal scales than have been previously possible.

## 1. Introduction

Viscoelastic coupling across the crust-mantle system is a fundamentally important component of earthquake physics, modulating time-dependent stress transfer, fault loading [e.g., *Roth*, 1988; *Tselentis and Drakopoulos*, 1990; *Freed and Lin*, 2001; *Zeng*, 2001; *Pollitz and Sacks*, 1997, 2002; *Pollitz et al.*, 1998, 2003; *Jónsson et al.*, 2003; *Bürgmann et al.*, 2002], and time-dependent surface deformation [e.g., *Hetland*, 2006; *Hearn et al.*, 2009; 2013]. Computational requirements have often limited the scope of viscoelastic model calculations to a few fixed rheological structures and model geometries over short (<10 yr) time periods at a given depth or specific location after a large earthquake [e.g., *Lorenzo-Martín et al., 2006*; *Pollitz and Sacks*, 2002; *Hearn et al.*, 2009]. The few studies that consider a large suite of viscosity structures [*Takeuchi and Fialko*, 2012; *DeVries et al.*, 2016a] require extensive computation time and resources (in the case of *DeVries et al.*, 2016a,b, 2.7 million CPU hours in total). These computational requirements make global-scale viscoelastic calculations across geometrically complex fault systems and large ranges of model parameters prohibitively expensive.

Here, we describe how large-scale viscoelastic calculations across geometrically complex fault systems with realistic slip histories can be radically accelerated by training a deep artificial neural network (ANN) to discover computationally efficient representations of a viscoelastic code. Artificial neural networks (ANNs) are machine learning algorithms that are widely used to detect patterns and trends and classify data. The purpose of ANN algorithms for this application can be best understood in the context of two common computational approaches in the earth sciences: forward modeling and state estimation methods (Table 1). Consider an arbitrary relationship $f$ between a set of inputs $x$ and a set out outputs $y$: $y = f(x)$. In simple terms, forward modeling approaches are aimed at solving for the outputs $y$, given $f$ and inputs $x$, while state estimation methods exist to solve for a set of inputs $x$ given $f$ and known outputs $y$. ANN algorithms, in contrast, are not aimed at solving for either inputs or outputs specifically, but rather at finding a new representation of $f$, the mapping between known inputs $x$, and known outputs $y$ (Table 1).

Viscoelastic codes take a set of model parameters, for example, the locations of the earthquake source and observations points, the source parameters (strike, dip, rake), viscosity structure and model geometry, as inputs, and perform computationally intensive calculations in order to ultimately output modeled displacements, velocities, strains, and stresses. Here, our goal is to use ANNs to find equivalent and compact representations of the mapping between viscoelastic code inputs and outputs. Once found, these efficient neural network representations may replace computationally intensive viscoelastic codes and allow large-scale viscoelastic calculations to be performed quickly and accurately.

Below, we first provide an overview of the viscoelastic calculations, followed by an explanation of the structure, development, and implementation of the ANNs used in our application. We then demonstrate the success of these machine learning methods (speedups of >50,000%) for both a simple elastic case and a complex and large-scale viscoelastic case. Finally, we discuss the implications of these methods for earthquake science and modeling efforts across the physical sciences.

## 2. Viscoelastic Code

Based on the semi-analytic framework developed by *Fukahata and Matsu'ura* [2005; 2006], we have implemented a three-dimensional viscoelastic earthquake cycle model based on a propagator matrix approach that enables the calculation of time-dependent displacements, velocities, strains, and stresses generated in response to fault slip in a coupled crust-mantle system. The distributed architecture of the code (97% efficiency on 1000 cores) facilitates the calculation of viscoelastic effects of earthquakes across realistic fault systems with highly complex geometries and slip histories. However, all told, these calculations may take thousands to millions of CPU-hours for large-scale applications [e.g. *DeVries et al.*, 2016a,b].

## 3. Artificial Neural Networks (ANNs)

Below, we briefly describe the basic set-up, structure, and workflow associated with developing these networks for our specific application; more extensive discussions of ANNs can be found in a number of machine learning textbooks [e.g., *Goodfellow et al.*, 2016; *Neilsen*, 2015]

### 3.1. Structure

Artificial neural networks consist of many interconnected neurons, or individual computational units (Figure 1a) [*Rosenblatt*, 1962]. In feedforward networks, neurons are organized into successive layers (Figure 1a). The first layer (Figure 1a, far left) corresponds to the inputs to the neural network, parameters such as the horizontal ($x$, $y$) position of the observations points with respect to the earthquake source, the time after the earthquake, the earthquake source depth, rake, and dip, and the viscosity structure. The last layer of neurons (Figure 1a, far right) corresponds to the predicted outputs of the neural network: displacements in the *x*-, *y*-, and *z*- directions, for example. The interior layers are referred to as 'hidden layers' (Figure 1a). More complex ANN structures are used for applications like image classification (convolutional ANNs [e.g., *LeCun et al.*, 1998a]) and speech recognition (recurrent ANNs [e.g., *Graves et al.*, 2013]), but for our application, feedforward networks perform extremely well.

Within ANNs, each neuron performs a simple operation: for input values $x_i$ from the previous layer (Figure 1b), each neuron outputs a single value $a = f(\sum w_i x_i + b)$ (Figure 1c, d). Activation functions $f$ are application-dependent [*LeCun et al.*, 1998b; *Glorot and Bengio*, 2010], but are usually chosen to be nonlinear or piecewise functions (e.g., the sigmoid or hyperbolic tangent functions; Figure 1c). The values of the weights $w_i$ and the bias $b$ are unique to each individual neuron (Figure 1c); these are the parameters that are iteratively adjusted during the training of the neural network. Neuron $j$ in layer $l$ accepts the outputs $a_{1...N}^{l-1}$ from all $N$ neurons in the previous layer *l-1* as its inputs, and outputs a single value $a_j^l = f(\sum w_i a_i^{l-1} + b)$ that is sent to every neuron in layer *l+1* (Figure 1b-d; *Neilsen* [2015]).

In practice, despite the universality principle for simple nets [e.g., *Cybenko*, 1989; *Hornik et al.*, 1989], researchers tend to choose the deepest (number of hidden layers) nets possible within their computational constraints.

*3.2. Training and Testing*

The workflow for developing ANNs consists of two phases: (1) training and validation and (2) testing. Both phases require inputs, in our case, sample inputs to the viscoelastic code, and associated code outputs. We build the training, validation, and testing data sets by running the viscoelastic code across a range of model parameters. All the data sets consist of vectors of input parameters (e.g., *x*- and *y*- locations relative to the horizontal position of the source) and output values from the viscoelastic code (e.g., the *x*-, *y*- and *z*-displacements associated with each set of input values). As described above, the neural network uses the input parameters as the first layer of neurons, and the last layer of neurons represent the predicted output values (Figure 1a).

During the training and validation phase, a cost (or loss) function C, the mean squared error, in this case, is used to evaluate the performance of the ANN by comparing the viscoelastic code outputs (from the training and validation data sets) to the values predicted by the ANN.

Optimization methods like stochastic gradient descent are used to iteratively solve for the different weights $w_i$ and biases $b$ for each neuron that allow the ANN to best approximate the training data outputs. At each iteration, a backpropagation algorithm estimates the partial derivatives of the cost function with respect to incremental changes of all the weights $w_i$ and biases $b$, to determine the gradient descent directions for the next iteration.

Validation data are not used in the optimization or backpropagation algorithms. Instead, the cost function evaluated over the validation data serves as an independent metric of the performance of the ANN; a validation data cost function that is increasing with each iteration would suggest that the net is overfitting, or memorizing, the training data, at the cost of generalizability to other inputs. Dealing with overfitting is not as significant a challenge for us as it is in other ANN applications, because there is no noise in the training data set in this case.

Finally, in the testing phase, we test the predictions of the ANN against the results from the viscoelastic code across different parameter values than the training set. This testing phase is necessary to evaluate how accurately the ANN can approximate the code outputs in a general sense.

*3.3. Implementation*

We use the Python toolkit Keras (https://github.com/fchollet/keras), which provides a high-level API to access the Theano (http://deeplearning.net/software/theano/) and TensorFlow (https://www.tensorflow.org/) deep learning libraries. For the examples shown here, we use Theano and an adaptive moment estimation (Adam) optimization method, a stochastic gradient descent algorithm that computes adaptive learning rates [*Kingma and Ba*, 2015].

**4. Results**

Below we demonstrate the success of these methods, in terms of both accuracy and acceleration relative to existing codes, for a simple elastic case [*Okada*, 1992] and for a complex and large-scale viscoelastic case.

*4.1. Elastic case*

We consider the simple case first: the displacements at the surface due to a strike-slip point source in an elastic half-space at a depth of 7.9 km [*Okada*, 1992]. In this case, the ANN is essentially acting as a two-dimensional interpolator: the inputs are *x*- and *y*-location, and the outputs are *x*-, *y*-, and *z*-displacements. A neural network trained on only 500 randomly distributed data points does not perform particularly well (mean residual of ~$3.4 \times 10^{-6}$ mm over a grid of 32,580 test points; Figure 2); however, as the density of training data is increased to over 40,000 points, the accuracy of the ANN solution rapidly improves (a mean residual of ~$1.9 \times 10^{-7}$ mm over a grid of 32,580 test points; Figure 2). The maximum absolute value of the x-displacements is ~0.5 mm, suggesting that this best-performing ANN is accurate to within 2 parts in 5 million on average. In addition, the ANN performs these calculations 165% faster than the explicit *Okada* [1992] calculation.

*4.2. Viscoelastic case*

This method also performs extremely well for a large-scale viscoelastic case involving 6 model parameters (Figure 3). In this more complex case, we fix the model geometry to be a 15-km thick elastic upper crust overlying a viscoelastic half-space with a transient Burgers rheology (shear moduli $\mu$ = 30 GPa in both layers) and consider only strike-slip sources. The input parameters to the artificial neural network are the (*x*, *y*) locations of the observations points at the surface relative to the source location, the depth of the earthquake source, the time after the earthquake, and the Maxwell and Kelvin viscosities $\eta_M$ and $\eta_K$ that characterize a Burgers rheology. The outputs predicted by the neural network are, as in the elastic case, *x*-, *y*-, and *z*-displacements.

For this large-scale test, the training data consists of 236,792,880 sets of inputs and outputs (~17 GB of training data in total): specifically, 10,609 (*x*, *y*) observation points at the surface, 20 source depths, 31 times, and 36 viscosity structures as inputs, and 3 components of displacement as outputs. In contrast to many ANN applications (e.g., voice recognition, image classification), we can exactly choose the training data in order to try to optimize the training of the ANNs. Training points are not regularly spaced in time, for example, because best results were obtained when the viscoelastic code solutions were sampled according to a quadratic sequence in time, in order to feed more information about the rapidly changing solution near *t* = 0 into the ANN.

We tested a range of deep ANN structures (10-40 neurons/layer and 2-10 hidden layers) and a number of different activation functions (hyperbolic tangent, sigmoid, and rectified linear units (ReLU), the last defined as *f(x)*=*x* for *x*>0 and *f(x)*=0 for *x*<0). The most successful nets (mean absolute residual of ~$2 \times 10^{-6}$ mm; Figure 3) incorporate alternating layers of hyperbolic tangent functions and ReLUs, with at least 10 neurons per layer and at least 6 hidden layers (e.g., Figure 4).

The performance of these ANNs, once trained, was evaluated comprehensively against the true viscoelastic solutions at 10,000 ($x$, $y$) locations for each of 13 source depths, 25 viscosity structures, and 81 times (Figure 3). The mean absolute residual, across all of these parameters, was ~$2 \times 10^{-6}$ mm; performance improves for deeper source depths but has no clear pattern as a function of time (Figure 3d,h,l). Perhaps most significantly, for this large-scale test (Figure 3), the viscoelastic code took 419,790 seconds to generate the true solution. The ANN prediction took 6366 seconds on a single CPU, a speedup of 6,600% compared to the viscoelastic code. Running the ANN prediction on a GPU further accelerates the same calculation to 763 seconds, a speedup of 55,018% over the viscoelastic code.

## 5. Discussion and Implications

Artificial neural networks can be used to discover new, compact, and accurate computational representations of viscoelastic physics (e.g., Figure 4). In other words, the structure, weights, and biases of trained ANNs are an alternative way to express the complex physics within existing computationally intensive viscoelastic codes. These new and efficient computational representations of viscoelastic physics are notable in a number of ways, listed below:

1. *Acceleration*: With trained ANNs, viscoelastic calculations can be accelerated by ~55,000%. Large-scale viscoelastic calculations [e.g., *DeVries et al.*, 2016a,b] that previously took over 2 million CPU hours could take only ~5 hours on an equivalent number of GPUs. With these trained ANNs, massive viscoelastic calculations, across very large spatial and temporal scales and ranges of model parameters, can be done in a matter of minutes or hours.

2. *Ease of use and portability*: To run forward predictions using the trained ANNs, only the matrices of weights $w_{ij}$, biases $b_j$, and activation functions are required. This information can generally be contained in a small (<50 KB) data file. Therefore, with only a ~100 line Python or Matlab script and a small data file, time-dependent viscoelastic deformation at any observation point and time, due to any source location, fault geometry, and rheological model, will be both easy and fast to calculate on any device with an installation of Matlab or Python.

3. *Unique physical insights*: From a philosophical perspective, these new ANN representations

of viscoelastic physics may lead to basic advances in the understanding of the underlying phenomenology and physics. For example, a visualization of one successful neural network (Figure 4) reveals that the weights connecting viscosity inputs (Maxwell and Kelvin viscosities $\eta_M$ and $\eta_K$) to the first hidden layer are on average at least a factor of 10 smaller in magnitude than all other connections. Therefore, in some sense, the viscosity structure in a viscoelastic calculation is an order of magnitude less influential over the full solution than other model parameters, such as the location of the source and observations points. Artificial neural networks have to the potential to illuminate countless fundamental physical insights along these lines.

4. *Implications for modeling across the physical sciences*: More broadly, this ANN method is exciting because it could be applied to virtually any type of physical model, accelerating model calculations and enabling larger parameter spaces and spatial and temporal scales to be explored in detail. These methods could therefore lead to novel insights across the physical sciences.

**6. Conclusions**

We enable viscoelastic calculations on a global scale across geometrically complex fault systems by training deep neural networks to accurately represent viscoelastic solutions. This method leads to accurate results with speedups of at least ~55,000%, allowing massive viscoelastic calculations to be done in minutes to hours, compared to the millions of CPU hours required to run large-scale viscoelastic calculations with existing codes. Once trained, artificial neural networks are new and compact representations of viscoelastic calculations, and as such may provide basic insights into viscoelastic physics and phenomenology. More broadly, this artificial neural network approach could be applied to virtually any physical model and contribute to the understanding of complicated physical behavior across many scientific fields.

## Acknowledgements

No data was used in producing the results in this manuscript. All data referred to in the discussion is properly cited and referred to in the reference list. The computations in this paper were run on the Odyssey cluster supported by the FAS Division of Science, Research Computing Group at Harvard University. This work was supported by Harvard University, the Southern California Earthquake Center (Contribution No. 6239) funded by NSF Cooperative Agreement EAR-1033462 and USGS Cooperative Agreement G12AC20038, and the Department of Energy Computational Science Graduate Fellowship Program of the Office of Science and National Nuclear Security Administration in the Department of Energy under contract DE-FG02-97ER25308.

| Computational approach | Goal |
| --- | --- |
| Forward models | Solve for output: $y = f(x)$ |
| State estimation methods | Solve for input: $y = f(x)$ |
| Artificial Neural Networks (ANNs) | Solve for model: $y = f(x)$ |

TABLE 1. Summary of goals of three computational approaches, to illustrate the purpose of ANNs for this application. Consider an arbitrary relationship $f$ between a set of inputs $x$ and a set out outputs $y$: $y = f(x)$. Forward modeling approaches are aimed at solving for the outputs $y$, given $f$ and inputs $x$, while state estimation methods exist to solve for a set of inputs $x$ given $f$ and known outputs $y$. ANN algorithms, in contrast, are not aimed at solving for either inputs or outputs specifically, but rather at finding a new representation of $f$, the mapping between inputs $x$, and outputs $y$.

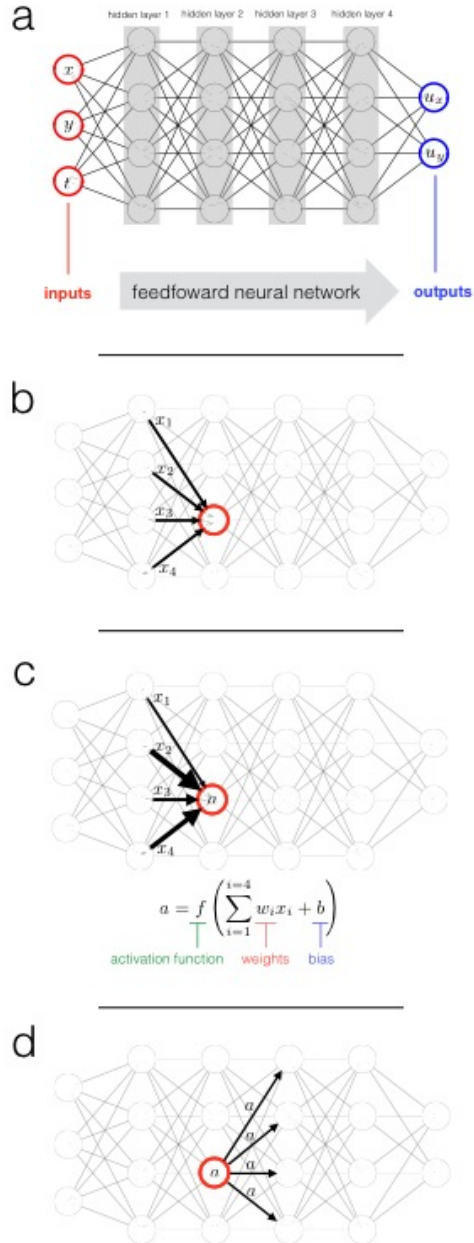

FIGURE 1. a) Structure and setup of an example artificial neural network with hidden layers, input layers, and output layers highlighted. Calculations move from left to right in the feedforward neural networks used here. b) Inputs to an example neuron from the previous layer. c) The calculation performed by the example neuron, weighting the inputs relative to one another, adding a bias, and applying an activation function, in order to calculation a value *a*, referred to as the activation of the neuron. d) The activation of the example neuron serves as one of the inputs to the next layer of neurons. Each neuron in the successive layers of the ANN are performing this same operation with different values of the tunable parameters *b* and $w_i$.

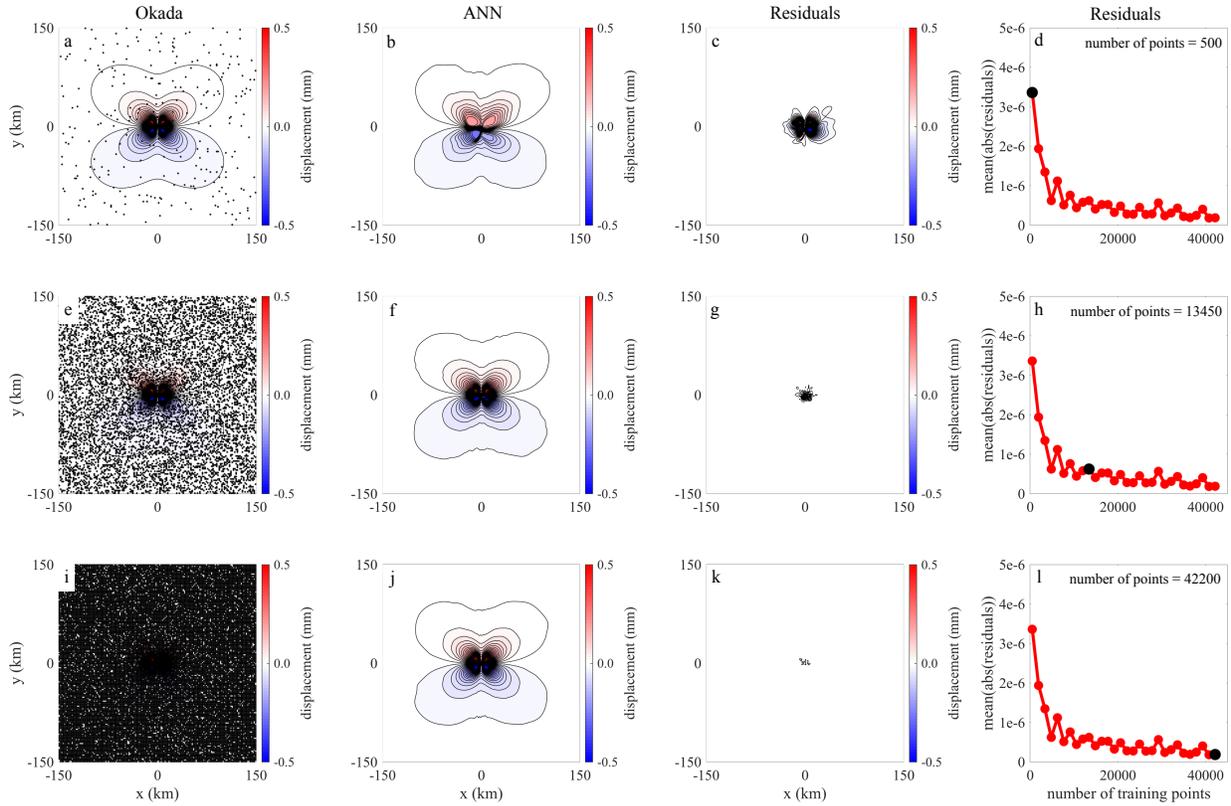

FIGURE 2. Performance of the artificial neural network for a simple elastic case. a) *Okada* [1992] solution for a point source at 7.9 km depth. Black dots indicate 500 randomly distributed training points used to train an ANN; predictions from this ANN are shown in (b); c) Residuals between the ANN prediction and the true *Okada* [1992] solution; d) Mean absolute residuals as a function of the number of training points for this elastic case; (e-h) Analogous to (a-d) for 13450 training points; (i-l) Analagous to (a-d) for 42200 training points.

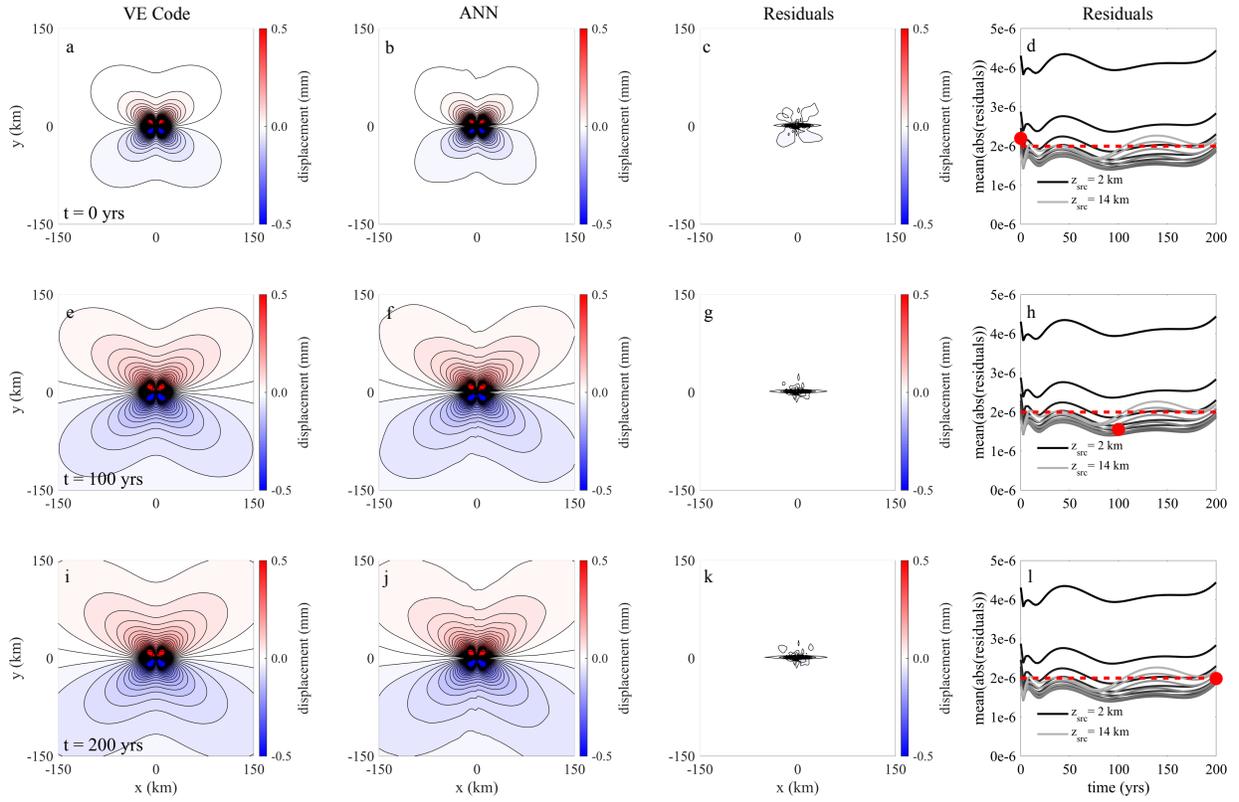

FIGURE 3. Performance of the artificial neural network for a complex and large-scale viscoelastic case. a) Viscoelastic code solution for a point source at 6 km depth for a Maxwell rheology of $10^{19}$ Pa s at $t = 0$ years after an earthquake; b) ANN prediction for comparison; c) Residuals between the true viscoelastic code solution and the ANN prediction; d) Mean absolute residuals as a function of time, averaged across all viscosity structures tested, for earthquake source depths between 2 and 14 km; e-h) Analogous to (a-d) for $t = 100$ years after an earthquake; i-l) Analagous to (a-d) for $t = 200$ years after an earthquake.

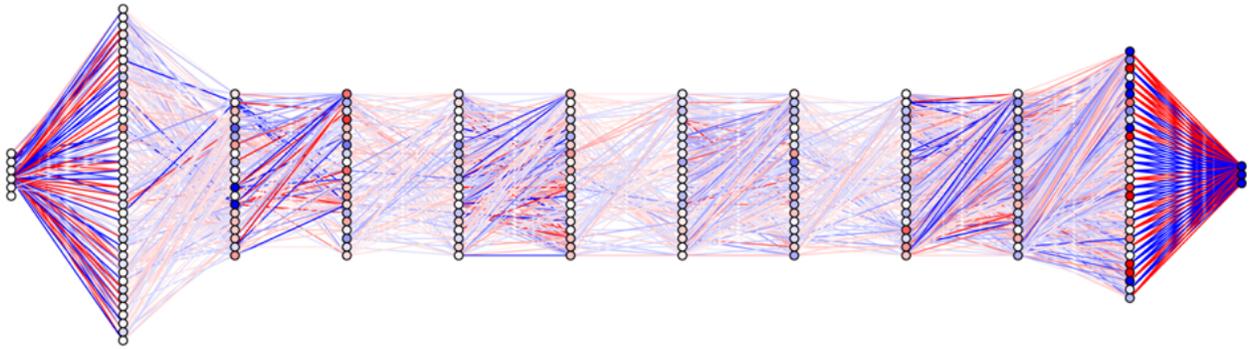

FIGURE 4. Visualization of the structure, weights, and biases of one successful trained ANN for the complex viscoelastic case (Figure 3). Circles represent individual neurons (as in Figure 1); the colors of the connections between neurons encode the magnitude and sign (red positive, blue negative) of each individual weight, while the colors of the neurons represent the sign and magnitude of the bias associated with each neuron. The information contained in this visualization (ANN structure, weights, biases) is, in some sense, a new and compact representation of viscoelastic physics.

Goodfellow, I., Y. Bengio, A. Courville (2016), Deep Learning, MIT Press, http://www.deeplearningbook.org.

Graves, A., A. Mohamed, G. Hinton (2013), Speech Recognition with Deep Recurrent Neural Networks, *2013 IEEE international conference on acoustics, speech and signal processing*, 6645-6649, Retrieved November 20, 2016, from the arXiv database (arXiv: 1303.5778v1).

Hearn, E. H., S. McClusky, S. Ergintav. and R.E. Reilinger (2009), Izmit earthquake postseismic deformation and dynamics of the North Anatolian Fault Zone, *J. Geophys. Res*., *114*(B8).

Hearn, E.H., C.T. Onishi, F.F. Pollitz, and W.R. Thatcher (2013), How do "ghost transients" from past earthquakes affect GPS slip rate estimates on southern California faults? *Geochem., Geophys., Geosys.*, *14*(4), 828-838.

Hetland, E. (2006), Models of interseismic deformation with an analytic framework for the inclusion of general linear viscoelastic rheologies. Ph.D. Thesis, Massachusetts Institute of Technology, 255 p.

Hornik, K., M. Stinchcombe, H. White (1989), Multilayer feedforward networks are universal approximators, *Neural Networks*, *2*(5), 359-366.

Jónsson, S., P. Segall, R. Pedersen, G. Bjornsson (2003), Post-earthquake ground movements correlated to pore-pressure transients, *Lett. Nature., 424*, 179-183.

Kingma, D. and J. Ba (2015), Adam: A Method for Stochastic Optimization, Conference paper at the 3rd International Conference for Learning Representations, San Diego, Retrieved October 25, 2016, from the arXiv database (arXiv: 1412.6980v8).

LeCun, Y., L. Bottou, Y. Bengio, P. Haffner (1998a), Gradient-based learning applied to document recognition, *Proceedings of the IEEE, 86*(11), 2278-2324.16